\documentclass[usenatbib,fleqn]{mn2e}

\newif\ifpdf
\ifx\pdfoutput\undefined
   \pdffalse              
\else
   \pdfoutput=1           
   \pdftrue
\fi
\ifpdf
  \usepackage[pdftex]{graphicx}
  \usepackage[bookmarks=true]{hyperref}
  \hypersetup{%
        bookmarksopen=true,
        colorlinks=true,
        pdfpagemode=UseOutlines,
        pdftitle={Merging of massive BHs due to ejection of bound stars},
        pdfsubject={},
        pdfauthor={Dr.~C.~Zier}
  }
  \usepackage{thumbpdf}
\else
  \usepackage[dvips]{graphicx}
\fi

\usepackage{xspace}
\usepackage{latexsym}
\usepackage{amssymb}
\usepackage{amsmath}
\usepackage{journals}

\newcommand{\eqn}[1]{Eq.~(\ref{#1})}
\newcommand{\eqb}[1]{(Eq.~[\ref{#1}])}
\newcommand{\fgr}[1]{Fig.~\ref{#1}}
\newcommand{\mrm}{\mathrm}

\newcommand{\X}{\textsf{X}\xspace}
\newcommand{\Z}{\textsf{Z}\xspace}
\newcommand{\BH}{BH\xspace}

\newcommand{\m}{-}
\newcommand{\D}{\displaystyle}

\def\beq{\begin{equation}}
\def\eeq{\end{equation}}
\def\beqa{\begin{eqnarray}}
\def\eeqa{\end{eqnarray}}

\title[Merging of a massive binary]{Merging of a massive binary due to
  ejection of bound stars}

\author[C. Zier]{C. Zier$^{1}$\thanks{E-mail: chzier@rri.res.in}\\
$^{1}$Raman Research Institute, Bangalore 560080, India}
\begin{document}

\date{Accepted Received}

\pagerange{\pageref{firstpage}--\pageref{lastpage}} \pubyear{2006}

\maketitle

\label{firstpage}

\begin{abstract}
From the results of numerical scattering experiments and simulations of a
massive black hole binary in spherically symmetric and shallow cores it has
been deduced that most likely the shrinking process stalls due to loss-cone
depletion before emission of gravitational radiation becomes important. Here
we follow a different approach and focus on the population of stars which is
bound to the binary and so far has not received much attention. With simple
assumptions which should not be sensitive to initial conditions we derive a
lower limit for the mass of stars which needs to be ejected by the binary in
order to coalesce. We also compute this mass in dependence on the steepness of
the density profile according to which the stars are distributed. Our results
are not as pessimistic as earlier conclusions and actually suggest that the
\BH{}s merge.
\end{abstract}


\section{Introduction}
\label{s_intro}
The existence of a massive black hole binary (BHB) is a natural consequence of
the assumptions that galaxies harbour a massive black hole (BH) in their
center and that galaxies merge with each other. The evolution of the binary
can be split up in roughly three phases: First both cores with the BHs in
their center spiral inwards to each other due to dynamical friction. When the
\BH{}s bind to each other and form a hard binary on the parsec scale, the
semimajor axis continues to decay due to slingshot ejection of stars.
Eventually, in the third phase, emission of gravitational radiation dominates
further shrinking until the \BH{}s coalesce \citep{begelman80}. It is still a
matter of debate whether the \BH{}s coalesce or the shrinking stalls before
entering the final phase because no stars are left to interact with the binary
(i.e. loss-cone depletion). However, the origin of \X- and \Z-shaped radio
galaxies is probably best explained by the coalescence of the two \BH{}s
\citep{rottmann01,zier02,gopal-krishna03,zier05} and the observed number of
these sources is in agreement with the merging rate derived for radio galaxies
\citep{merritt02}. Numerical scattering experiments showed that the \BH{}s can
merge on scales of $\sim 10^{8\text{--}9}\,\mrm{yr}$ if the loss-cone is
always full \citep{quinlan96,quinlan97,zier01,milos01}. However, for a flat
core most authors estimate that the loss-cone gets depleted long before the
binary can enter the third phase. Further hardening then depends on the rate
at which the loss-cone is refilled, most probably by two-body scattering of
stars. This implicates a time scale for the separation of the \BH{}s to shrink
to a distance where gravitational radiation becomes dominant which might
exceed a Hubble time so that the binary basically stalls. The conclusions are
essentially the same: though the results show that the BHs coalesce it is
argued that in real galaxies the loss-cone is depleted before gravitational
radiation dominates further shrinking. These conclusions are based on the
assumption that at the time the binary becomes hard the density profile of the
cusp is flat (i.e. with a power-law index of at most 2, usually 1 or even
0). This seems to have been confirmed by recent simulations of a binary in a
constant density core \citep{berczik05}. As the authors say themselves the
choice for the mass of the binary is unrealistically large compared to that of
the galaxy and the initial conditions are quite unlikely, because the \BH{}s
were introduced symmetrically about the center of the galaxy instead of
arriving there due to the evolution of the merger.

The initial conditions in these numerical experiments are idealised and cover
only a small fraction of the parameter space. They are based on spherically
symmetric profiles which have been derived from observations of elliptical
galaxies. These profiles correspond to the central cluster before or after the
merger when mass has already been ejected from the loss-cone and shifted from
smaller to larger radii, resulting in flatter profiles. However, they do not
match the central profiles \emph{during} the merger. Also ignored are
individual spins of both galaxies, which might stabilize the cluster against
tidal disruption by the other \BH beyond a hard binary, depending on the
magnitude of the spins and their orientation relative to each other and
relative to the orbital angular momentum of the merging galaxies. These
parameters have a strong influence on the merger itself and the morphology of
the remnant, as has been shown by \citet{toomre72}. When galaxies collide
energy will be dissipated and angular momentum redistributed with some
fractions compensating each other. Large amounts of mass will move on highly
eccentric orbits \citep{rauch96} in a potential that is strongly
non-spherically symmetric so that the angular momentum of a single particle is
not conserved and matter with low angular momentum piles up in the center. The
density of both cores will be increased considerably before they merge and the
binary becomes hard \citep{barnes96}. Each of the \BH{}s will carry a stellar
cusp as massive as the \BH and therefore a stellar mass comparable to that of
the binary will be concentrated in the central core once the cusps
merge. Hence we expect that at the time the \BH{}s form a hard binary, the
surrounding density distribution will be much more compact with a steeper
profile than in non-interacting galaxies, i.e. unlike the initial conditions
used so far in numerical experiments. It is the fraction of this cusp which is
bound to the binary to which our analyses applies.

Here we will study in a simple approach how compact such a cusp has to be and
wich profile is required so that the binary can shrink to the third phase and
the \BH{}s eventually coalesce. Whereas numerical experiments were more
concerned with unbound stars scattered off the binary, we will focus on stars
which are bound by the \BH{}s. Therefore our results will be less sensitive to
initial conditions.


\section{Preliminaries}
Being interested in the 2nd phase of the merger we assume that the BHs have
bound to each other and are moving on Keplerian orbits. The origin is the
center of mass of the binary and we define the mass ratio $q\equiv m_2/m_1
\leq 1$. The total and reduced mass are $M_{12} = m_1 + m_2$ and $\mu = m_1
m_2/M_{12}$, respectively. With $a$ we denote the semimajor axis of the
binary, i.e. of the orbit of the reduced mass. For circular orbits this is
equal to the separation of the BHs. Thus the energy of the binary can be
written as
\begin{equation}
E_\mrm{bin} = -\frac{GM_{12}\mu}{2a},
\label{eq_e-bin}
\end{equation}
and the relative velocity between the BHs is
\begin{equation}
v_\mu = \sqrt{\frac{GM_{12}}{a}},
\label{eq_v-bin}
\end{equation}
which corresponds to the velocity of the reduced mass if it moves on circular
orbits. The definition for the semimajor axis $a_\mrm{h}$ where the binary
becomes hard is not unique and changes in the literature
\cite[e.g.][]{heggie75,hills75a,hut83,quinlan96,milos01}. Some definitions are
derived from the results of numerical experiments and are more
phenomenological.  Usually they are all based on comparing the velocity
dispersion of the cluster (which itself is not so easy to define in an ongoing
merger when the core is far from being relaxed) with the velocity of a
component of the binary. For major mergers, i.e. large $q$, they all yield
similar results for $a_\mrm{h}$, which is found to be on the parsec scale for
$M_1\simeq 10^8\,M_\odot$.

We think the transition from the first to the second phase is best defined by
the distance between the \BH{}s when the stars are moving in the potential of
both \BH{}s, not just one, so that the binary and the star can be treated as a
restricted three body problem. This transition from dynamical friction to
slingshot ejection as dominating processes for the decay of the binary is
smooth. Because we do not know an exact definition for $a_\mrm{h}$ we instead
scale it to the semimajor axis $a_\mrm{g}$ at the end of the second phase,
which is more clearly defined. An upper limit for this transition is if it
still takes a Hubble time for the BHs to merge completely from $a=a_\mrm{g}$
due to emission of gravitational radiation. For circular orbits this is
\citep{peters64}
\begin{equation}
\begin{split}
a_\mrm{g} & = \left[\frac{256}{5}\, \frac{G^3\mu M_{12}^2}{c^5}\, t_\mrm{g}
  \right]^{1/4}\\
& \approx \frac{1}{15} \left(\frac{M_1}{10^8 M_\odot}\right)^{3/4}
  \left(\frac{t_\mrm{g}}{10^{10}\,\mrm{yr}}\right)^{1/4}
  [q(1+q)]^{\frac{1}{4}}\,\mrm{pc}.
\end{split}
\label{eq_a-g}
\end{equation}
In previous publications the ratio of the semimajor axes where the transitions
between the phases occur, i.e. $\eta\equiv a_\mrm{h}/a_\mrm{g}$, has been
found to be in the range of $20\,\text{--}\,100$. Applied to \eqn{eq_a-g} this
yields $a_\mrm{h}$ to be in the range $1\,\text{--}\,7\,\mrm{pc}$, in
agreement with $a_\mrm{h}$ being on the parsec scale. Because this ratio does
not seem to depend sensitively on the initial conditions and is quite robust
we will use it in the following to compute the location of $a_\mrm{h}$. Note
that $a_\mrm{h}$ and $a_\mrm{g}$ scale differently with the mass of the \BH{}s
and the velocity dispersion of the cluster. Consequently the ratio $\eta$ will
be a function of these quantities. Because we are considering a range of fixed
values for $\eta$, this does not affect our following analysis and might
become important only if systems with very different \BH masses and velocity
dispersions are considered.

For bound stars it makes only a small difference whether the stellar
distribution and potential is spherical, axisymmetric or triaxial. For
simplicity we consider a spherical density distribution which in the inner
region follows a power-law, $\rho = \rho_0 (r/r_0)^{-\gamma}$. With the total
mass of this cluster ($M_\mrm{c}$) being distributed between the inner and the
cluster radius, $r_\mrm{i}$ and $r_\mrm{c}$ respectively, the mass
distribution is
\begin{equation}
M(r) = M_\mrm{c} \begin{cases} \frac{\D r^{3-\gamma} -
  r_\mrm{i}^{3-\gamma}}{\D r_\mrm{c}^{3-\gamma} - r_\mrm{i}^{3-\gamma}}, &
  \gamma \neq 3\\ \frac{\D \ln(r/r_\mrm{i})}{\D \ln(r_\mrm{c}/r_\mrm{i})}, &
  \gamma = 3.
\end{cases}
\label{eq_m}
\end{equation}
To avoid a singularity at $r=0$ we cut off the distribution at the inner
radius $r_\mrm{i}$. The circular velocity is $v_\mrm{circ}^2 = r\,d\Phi/dr = G
M(r)/r$. A star with this velocity is bound to a BH with mass $M_{12}$ at the
center of the cluster if the velocity is less than the escape velocity
$v_\mrm{esc} = \sqrt{2GM_{12}/r}$. Using the circular velocity as the typical
velocity (i.e. the velocity at the maximum of an isothermal sphere with a
Maxwellian velocity distribution, same as $\sqrt{2}$-times the velocity
dispersion of this distribution) we obtain from $v_\mrm{circ}\leq v_\mrm{esc}$
the relation $M(r)\leq 2 M_{12}$. Thus a mass of about $2M_{12}$ of the
cluster is bound to the binary. We expect a large fraction of this mass to be
in the loss-cone.


\section{Required mass and its distribution}
In this section we will derive limits for the mass which is required to be
ejected and how steep the mass distribution has to be so that the \BH{}s can
merge. The inner region where the stars are bound to the binary is dominated
by the potential of the \BH{}s. Approximating the potential of the binary to
first order with a point potential of the mass $M_{12}$ located at the
cluster's center introduces only minor deviations whith a maximum of a factor
less than $2$ for $q=1$. Before a star becomes ejected via the slingshot
mechanism its binding energy is about
\begin{equation}
E_{\ast,\mrm{i}} = -(1-\epsilon) \frac{GM_{12} m_\ast}{2r_\m},
\label{eq_e-star-i}
\end{equation}
where $r_\m$ denotes the pericenter and $\epsilon < 1$ the eccentricity of the
orbit. In this expression we neglected the potential of the cluster itself,
whose mass amounts up to twice the mass of the binary, as we have shown in the
previous section. The influence of the potential of the cluster could be
explored in self-consistent calculations using a potential generated by the
cluster until its mass drops below that of the binary. We do not expect the
cluster potential to change the basic results obtained in the present work and
leave these calculations to a subsequent paper. The final energy of the star
after its ejection, $E_{\ast,\mrm{f}}$, is zero or more. Scaling it to the
initial energy with $\epsilon = 0$ we can write it as $E_{\ast,\mrm{f}} =
\kappa\, G M_{12} m_\ast/2r_\m$, with $\kappa$ being the scaling
factor. Independent of the density profile \citet{quinlan96} finds that the
dominant contribution to the hardening of the binary comes from stars whose
closest approach to both BHs is about the semimajor axis $a$. Hence we replace
$r_\m$ with $a$ so that the energy change of the star $E_{\ast,\mrm{f}} -
E_{\ast,\mrm{i}}$ can be written as
\begin{equation}
\Delta E_\ast = (1-\epsilon +\kappa) \frac{G M_{12}m_\ast}{2a} \equiv
k\,\frac{G M_{12}m_\ast}{2a},
\label{eq_de-star}
\end{equation}
with $k$ being defined by the last equality. From scattering experiments it
follows that $\kappa\approx (3/2)^2 m_2/M_{12}$ \citep{quinlan96} or $\sim
\mu/M_{12}$ \citep{saslaw74}. Thus for circular orbits ($\epsilon = 0$) we
find $k$ in the range $1\lesssim k \lesssim 2$, or as large as $3.2$ according
to \citet{yu02}. Even for highly eccentric orbits ($\epsilon\sim 0.7$) the
energy change will be of the order of one.


\subsection{The ejected mass}
When a star is ejected it extracts the amount of energy given in
\eqn{eq_de-star} from the binary. In the limit $m_\ast \ll m_2$ we can replace
$m_\ast$ with $dm$ and using \eqn{eq_e-bin} we can write $dE_\mrm{bin} =
\Delta E_\ast$ as
\begin{equation}
\frac{da}{a} = -k \frac{dm}{\mu},
\label{eq_da}
\end{equation}
giving an expression for the shrinking of the binary $da$ due to the
ejection of a mass $dm$. Stars are ejected all the time as long as the
loss-cone is not depleted and so the binary hardens continuously. Integration
from $a_\mrm{g}$ to $a_\mrm{h}$ yields the mass that has to be ejected by the
binary in order to enter the last phase and we obtain
\begin{equation}
m_\mrm{ej} = \frac{\mu}{k} \ln \frac{a_\mrm{h}}{a_\mrm{g}}.
\label{eq_mej}
\end{equation}
This is only slightly more than $M_{12}$ for $q=k=1$ and $\eta = 100$ and
therefore less than the mass bound to the binary. In deriving this expression
we assumed that the stars are moving on orbits with pericenters which are
equal to the current semimajor axis of the binary. However, stars interior to
the binary will at least be disturbed, if not ejected, once the binary has
become hard. These stars are more tightly bound in the potential of the \BH{}s
and have to gain energy on the expense of the binary in order to be shifted to
orbits with $a$ as pericenter. This issue will be explored in detail in the
next section which shows that the mass obtained in \eqn{eq_mej} is sufficient
for a merger if it is distributed with a power-law index $\gamma=3$.

To define the mass ejection rate \citet{quinlan96} introduced a similar
expression as \eqn{eq_da} and used as mass scaling factor $M_{12}$ without
further justification. This is plausible for large $q$ because in this case it
can be expected that the binary has to eject a mass of about its own. This was
confirmed by his results which are essentially in agreement with
ours. However, in the limit $m_2\rightarrow m_\ast$ it does not seem to be
plausible that a total mass of about $M_{12}$ is required. The energy of the
binary is that of the reduced mass orbiting in the fixed potential of the
point mass $M_{12}$ with an angular momentum of $L_\mrm{bin} = \mu a
v_\mu$. Therefore a mass of about $\mu$, as derived in the above equation,
seems to be more natural. In order to explain mass deficits in galaxy cores
\citet{milos02} find that a scaling factor of $m_2$ instead of $M_{12}$
matches the ejected mass better. This dependency on $q$ is very similar to
that in our expression, supporting our approach of focussing on the bound
population of stars. Because we did not make use of any further assumptions in
the derivation of \eqn{eq_mej} it should best describe the true ejected mass.


\subsection{The distribution of the ejected mass}
According to \citet{milos01} the results of their simulations indicate that
soon after the binary becomes hard the loss-cone becomes depleted in little
more than the local crossing time of the cusp. If we assume that all the mass
of \eqn{eq_mej} is ejected instantaneously at $a_\mrm{h}$, i.e. that the
density is distributed according to a Dirac-delta distribution that is
non-vanishing at $r=a_\mrm{h}$, the energy change due to ejection of this mass
is $\Delta E_\mrm{ej} = (GM_{12}\mu/2 a_\mrm{h}) \ln(a_\mrm{h}/a_\mrm{g})$
(Eq.~[\ref{eq_de-star}]). Equating this with the change in energy of the
binary in \eqn{eq_e-bin} if the semimajor axis shrinks from $a_\mrm{h}$ to the
final semimajor axis $a_\mrm{f}$
\begin{equation}
\Delta E_\mrm{bin} = \frac{GM_{12}\mu}{2 a_\mrm{h}}
\left(\frac{a_\mrm{h}}{a_\mrm{f}} -1 \right)
\label{eq_de-bin}
\end{equation}
and solving for their ratio we obtain
\begin{equation}
\frac{a_\mrm{h}}{a_\mrm{f}} = 1+\ln(a_\mrm{h}/a_\mrm{g}),
\label{eq_af-delta}
\end{equation}
which is only about $5.6,\, 4.9$ and $4.0$ for $a_\mrm{h}/a_\mrm{g} =
100,\,50$ and $20$ respectively. However, stars at radii smaller than
$a_\mrm{h}$ are also bound deeper in the potential of the binary. Before the
above derivation for $a_\mrm{f}$ applies the BHs have to transfer sufficient
energy to these stars so that they move on orbits with a radius of
$a_\mrm{h}$. In the point potential the binding energy of a mass element $dm =
4\pi r^2 \rho(r)\,dr$ in a distance $r$ is $dE = GM_{12} dm/2r$. The
integration over the cusp's profile between $a_\mrm{g}$ and $a_\mrm{h}$ gives
the binding energy of the cusp in this potential and corresponds to the energy
which the binary loses when it ejects these stars. This negelects the
potential of the cusp itself, and because the final energies of the stars are
probably more than zero this is actually a lower limit for the energy loss of
the binary. Using the density profile $\rho = \rho_0 (r/r_0)^{-\gamma}$, and
with the help of \eqn{eq_m}, the integration yields
\begin{equation}
\Delta E = \frac{GM_{21}M_\mrm{c}}{2}
  \begin{cases} \frac{\D 1}{\D r_\mrm{c}-r_\mrm{i}}
    \ln(a_\mrm{h}/a_\mrm{g}) & \gamma = 2\\ \left(\frac{\D 1}{\D a_\mrm{g}} -
    \frac{\D 1}{\D a_\mrm{h}}\right) \left(\ln\frac{\D r_\mrm{c}}{\D
    r_\mrm{i}}\right)^{-1} & \gamma = 3\\ \frac{\D 3-\gamma}{\D 2-\gamma}\;
    \frac{\D a_\mrm{h}^{2-\gamma} - a_\mrm{g}^{2-\gamma}}{\D
    r_\mrm{c}^{3-\gamma} - r_\mrm{i}^{3-\gamma}} & \text{else}.
\end{cases}
\label{eq_e-cusp}
\end{equation}
We assume that the lower limit for the mass which is required to allow the BHs
to merge, found in \eqn{eq_mej}, is the cluster mass $M_\mrm{c}$ which is
distributed between $r_\mrm{i}$ and $r_\mrm{c}$ according to the above power
law. If the binary ejects the stars from the region $a_\mrm{g}\leq r \leq
a_\mrm{h}$ we can compute the final semimajor axis of the binary in the same
way as above, i.e. equating Eqs.~(\ref{eq_de-bin}) and (\ref{eq_e-cusp}), and
find:
\begin{figure}
\begin{center}
\ifpdf \includegraphics[width=86mm]{f_afag.pdf} \else
  \includegraphics[width=86mm]{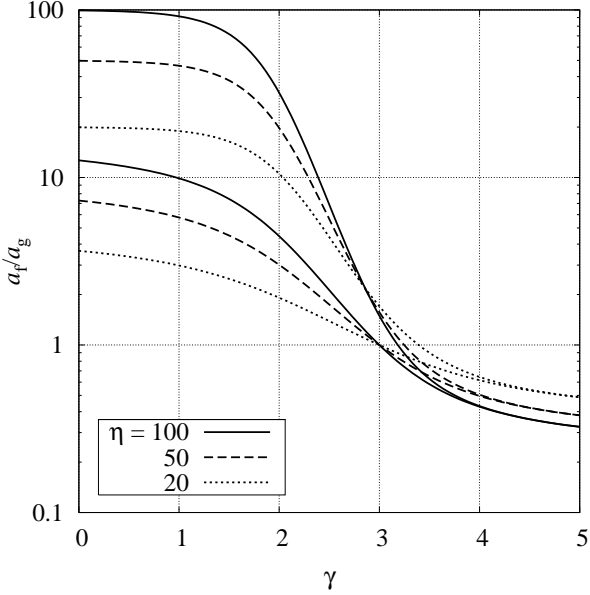} \fi
\caption[]{The final semimajor axis in units of $a_\mrm{g}$ as function of the
  exponent $\gamma$. Solid, dashed and dotted lines show the distribution for
  different ratios $\eta=a_\mrm{h}/a_\mrm{g}$. For $\gamma\lesssim 4$ they
  split up in a lower ($\lambda = 1$) and upper ($\lambda = 10$) branch.}
\label{f_af} 
\end{center}
\end{figure}
\begin{equation}
\frac{a_\mrm{h}}{a_\mrm{f}} = \begin{cases} 1 + \frac{\D 1}{\D
    k\lambda}\frac{\D \zeta}{\D \zeta-1} (\ln\eta)^2 & \gamma = 2\\ 1 +
    \frac{\D \eta- 1}{\D k}\frac{\D \ln\eta}{\D \ln\zeta} & \gamma = 3\\ 1+
    \frac{\D \lambda^{\gamma -3}}{\D k} \frac{\D 3-\gamma}{\D 2-\gamma}\,
    \frac{\D 1 - \eta^{\gamma -2}}{\D 1 - \zeta^{\gamma -3}} \ln\eta &
    \text{else.}
\end{cases}
\label{eq_af-potential}
\end{equation}
Here we have defined the following ratios: $\eta = a_\mrm{h}/a_\mrm{g} > 1$ as
before, $\zeta\equiv r_\mrm{c}/r_\mrm{i}> 1$ and $\lambda\equiv
r_\mrm{c}/a_\mrm{h}\geq 1$. If we allow $r_\mrm{i} \leq a_\mrm{g}$, so that
the cluster extends to radii smaller than $a_\mrm{g}$, the fraction of mass
between these two radii is not included in the ejected mass since we used
$a_\mrm{g}$ as lower limit in the integration \eqb{eq_e-cusp}. The steeper the
distribution, the larger is this mass fraction, and hence the less the binary
will shrink so that $a_\mrm{f}/a_\mrm{g}$ increases with $\gamma$, if $\gamma$
exceeds a certain value. However, these solutions are unphysical and will not
be considered.  Therefore, setting $r_\mrm{i} = a_\mrm{g}$, we can substitute
$\zeta = \lambda\eta$. In \fgr{f_af} we plotted the result for $k=1$ as ratio
$a_\mrm{f}/a_\mrm{g} = \eta\, a_\mrm{f}/a_\mrm{h}$ in dependence on the
exponent $\gamma$. As expected, the steeper the cusp the closer the binary
shrinks to $a_\mrm{g}$. In the range $\gamma \lesssim 4$ the solutions split
into a lower ($\lambda = 1$) and upper ($\lambda = 10$) branch. In case of
$\lambda = 1$ the mass of \eqn{eq_mej} is distributed between $a_\mrm{g}$ and
$a_\mrm{h}$, which coincide with $r_\mrm{i}$ and $r_\mrm{c}$ respectively. For
a uniform density distribution ($\gamma = 0$) the final semimajor axis is
smaller by a factor of about $1.4$ than for the case of the Dirac-delta
distribution which yields $a_\mrm{f}/a_\mrm{g} \approx 17.8,\,10.2$ and $5.0$
for $\eta = 100,\,50$ and $20$ respectively, see \eqn{eq_af-delta}. The
decrease of the final semimajor axis with increasing $\gamma$ is steeper for
large $\eta$. If we write in \eqn{eq_da} $dm = 4\pi r^2 \rho(r) dr$ and solve
for the density, we obtain $\rho(r) = \mu/4\pi k r^3$, i.e. the power-law for
which $m_\mrm{ej}$ is just enough to allow the \BH{}s to merge
($a_\mrm{f}/a_\mrm{g} = 1$ at $\gamma =3$). For a distribution as steep or
steeper than $\rho \propto r^{-3}$ the ejected mass is bound deeply enough in
the potential of the binary prior to its ejection for all ratios $\eta$ in
order to allow the BHs to shrink to the distance where gravitational radiation
begins to dominate the further decay. Hence for sufficiently steep and compact
distributions the \BH{}s coalesce.

If $\lambda = 10$ the mass $m_\mrm{ej}$ is distributed in a larger sphere with
an outer radius 10 times as large as $a_\mrm{h}$. The larger $\lambda$ is the
closer the ratio $a_\mrm{f}/a_\mrm{g}$ approaches $\eta$ at $\gamma = 0$
because less mass is available for ejection at radii $\le a_\mrm{h}$, and the
less the binary will shrink (upper branches in \fgr{f_af}). The dependency on
the exponent is basically the same as for $\lambda = 1$ and for sufficiently
steep distributions the ejected mass fraction removes sufficient energy so
that the \BH{}s can coalesce. At $\gamma = 3$ we can approximate
$a_\mrm{f}/a_\mrm{g}$ with $1+ \ln \lambda/\ln\eta$, a function which
increases slowly with $\lambda$. Even for the cluster extending as far as
$r_\mrm{c} = \eta a_\mrm{h}$ only about twice as much mass is needed for the
\BH{}s to coalesce as compared to $\lambda = 1$.
\begin{figure}
\begin{center}
\ifpdf \includegraphics[width=86mm]{f_mmej.pdf} \else
  \includegraphics[width=86mm]{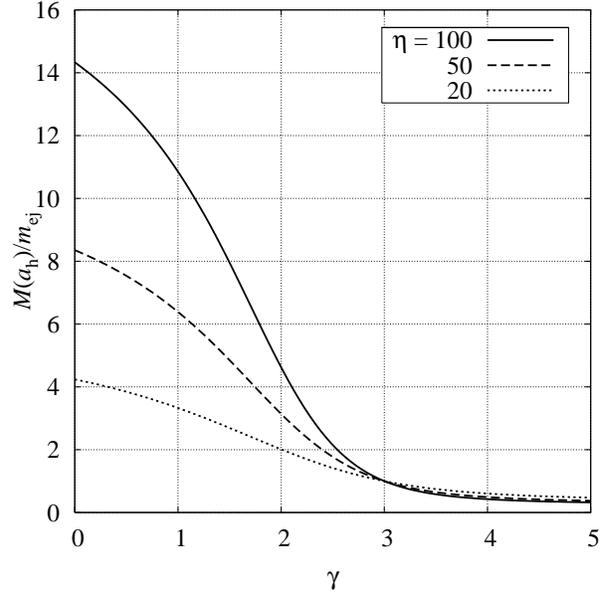} \fi
\caption[]{The mass distributed between $a_\mrm{g}$ and $a_\mrm{h}$ which is
  required to allow the \BH{}s to shrink to $a_\mrm{g}$ in units of
  $m_\mrm{ej}$ as function of the exponent $\gamma$. ($k = 1$).}
\label{f_m-mej} 
\end{center}
\end{figure}

To obtain the mass required for coalescence of the \BH{}s for other exponents
than $\gamma = 3$, where exactly $m_\mrm{ej}$ is needed (see \fgr{f_af}), we
can proceed as follows. In \eqn{eq_af-potential} we assume the cluster mass to
be distributed between $a_\mrm{g}$ and $a_\mrm{h}$, i.e. $\lambda = 1$ and
$\zeta = \eta$. Scaling the required mass to $m_\mrm{ej}$ with a factor $\chi
\equiv M(a_\mrm{h})/m_\mrm{ej}$ we just need to replace $1/k$ with $\chi /k$
in \eqn{eq_af-potential} and solve $a_\mrm{h}/a_\mrm{f} = \eta$ for
$\chi$. This yields
\begin{equation}
\frac{M(a_\mrm{h})}{m_\mrm{ej}} = k\, \frac{2-\gamma}{3-\gamma} \,
\frac{1-\eta^{\gamma -3}}{1-\eta^{\gamma -2}}\, \frac{\eta -1}{\ln\eta}.
\label{eq_m-mej}
\end{equation}
Plotting this ratio as function of the exponent for $k=1$ again shows that for
flat distributions ($\gamma\lesssim 2$) much more mass is required than for
steeper cusps, at least for $\eta\gtrsim 50$ (\fgr{f_m-mej}). For $\gamma
\approx 2.5$ this is less than $2\,m_\mrm{ej}$ for all $\eta$, confirming our
conclusion that the binary is likely to merge in compact cusps which are
steeper than $\gamma =2$.

Steep density distributions with $\gamma\gtrsim 2$ have not been explored in
numerical simulations or scattering experiments which used cusps at most as
steep as $\gamma = 2$. Considering the population of stars which is bound to
the binary we obtain quite different results which show that the BHs can
actually merge if the cusp is steep enough.


\section{Discussion and Conclusions}
Although numerical simulations showed that the \BH{}s in a hard binary
coalesce after about $10^{7\text{--}8} \,\mrm{yr}$ if the loss-cone is full
\citep[e.g.,][]{quinlan96,milos01,zier01} most authors argued that the
loss-cone becomes depleted and the binary stalls. This conclusion is based on
the assumption of a spherically symmetric shallow density profile and has been
confirmed recently by \citet{berczik05}, using initial conditions which are in
favour of a stalled binary. The choice of a flat central density distribution
is based on profiles derived from observations of elliptical galaxies,
i.e. galaxies before or (more probably) after a merger, when mass has been
redistributed from the inner to the outer parts of the cluster, resulting in a
flatter profile. However, \emph{during} the merger the profile might be quite
different and much steeper. Simulations by \citet{toomre72} support this
conjecture and \citet{barnes96} showed that both cores accumulate a big amount
of mass before they merge and the binary becomes hard. Each of the \BH{}s will
carry a stellar cusp with a mass of about its own.

Hence in this Letter we suggest that by the time the \BH{}s become hard, mass
and angular momentum have been redistributed and energy dissipated so that the
central cusp is more massive and has a steeper density profile than that
deduced from observations. Naturally this is a transient distribution and
unlikely to be observed because of its short lifetime. Because a mass of about
$2\,M_{12}$ will bind to the binary, a large fraction of which will probably
be contained in the loss-cone, we focused on this population, unlike numerical
experiments which so far have focussed on scattered unbound stars. By
computing the binding energy of a stellar distribution in the potential of a
binary we could show that the ejection of this population can play a decisive
role for a successful merger. From the energy extracted from the binary by
these stars due to sling-shot ejection we derived a new expression for the
mass which is required to be ejected so that the \BH{}s coalesce
\eqb{eq_mej}. While previous work assumed this mass to be proportional to the
mass of the binary $M_{12}$ due to the choice of the scaling factor, we find
it to be proportional to the reduced mass. This seems to be more plausible,
especially in the limit of a small secondary \BH, and is also in good
agreement with fits to the mass deficit derived from observations
\citep{milos02}. Because the mass bound to the binary is more than twice as
much as the mass required for coalescence we conclude that the fraction of
bound stars in the loss-cone contributes at least a significant fraction to
the coalescence and might be enough to enable merging until $a = a_\mrm{g}$ on
its own. We assumed that this mass is moving on circular orbits and
approximated the potential of the binary by that of a point mass. In
Figs.~\ref{f_af} and \ref{f_m-mej} we showed that the ejection of mass, which
is distributed in the potential of the binary according to a flat profile
($\gamma \lesssim 2$) as in previous papers, does not allow the \BH{}s to
merge. Either the binary stalls or a very large amount of mass is
required. However, these figures also show that for steeper distributions the
ejected mass is indeed sufficient to remove enough energy from the binary so
that the \BH{}s can coalesce.

If we allow the stars to move on eccentric orbits using their former radius as
pericenter, the binding energy will decrease while the required ejected mass
will increase. However, this will be more than compensated by the
non-vanishing velocity of ejected stars at infinity, which we assumed to be
0. Comparing \eqn{eq_de-star} with \citet{yu02} we find $k=3.2$, including
elliptic orbits. Elliptic orbits allow the matter to be distributed less
steeply than what is suggested by our results and therefore relax the
conditions for a complete merger. We also assumed the \BH{}s to move on
circular orbits. If they are elliptic gravitational radiation dominates the
shrinking earlier, shifting $a_\mrm{g}$ to larger radii and consequently
reducing the mass required for ejection. Therefore our results in fact are
quite conservative and we conclude that the \BH{}s coalesce in most cases,
contrary to previous results which were based on flat cores. Our analyses can
also be applied to dark matter. If there is not sufficient baryonic matter for
a merger and the \BH{}s still merge, our approach allows some conclusions to
be drawn about the central distribution of dark matter.

Simulations of merging \BH{}s in a triaxial potential
\citep{yu02,holley-bockelmann02,holley-bockelmann06,berczik06} strongly
support the conclusion that the \BH{}s merge completely in less than a Hubble
time. Due to the triaxiality of the potential, the central regions remain
populated. The flow of orbits through phase space into the loss-cone is large
enough to keep it always filled. An unfilled loss-cone is only found for
shallow spherically symmetric cores \citep{yu02}. This is an idealized
distribution which, if at all, applies only to the cluster after the merger
has been completed, but not to an ongoing merger. Thus, additionally to the
population bound to the binary that we were investigating here, simulations of
a triaxial potential show that the loss-cone remains filled during the merger,
supplying the binary with an almost unlimited amount of stars that can be
scattered off. Even in the case that the bound stars alone are not enough to
allow the \BH{}s to merge, together with the scattered stars from the
loss-cone a stalled binary should be most unlikely.

In a future paper (in preparation) we will discuss this approach and its
context in more detail.

\section*{Acknowledgements}
I would like to acknowledge the generous support and very kind hospitality I
experienced at Raman Research Institute. I also would like to thank Wolfram
Kr\"ulls for his valuable comments to improve this manuscript. I am also
grateful to the anonymous referee for helpful advice and comments.

\bibliography{$HOME/LATEX/refs}

\begin{thebibliography}{}

\bibitem[\protect\citeauthoryear{{Barnes} \& {Hernquist}}{{Barnes} \&
  {Hernquist}}{1996}]{barnes96}
{Barnes} J.~E.,  {Hernquist} L.,  1996, \apj, 471, 115

\bibitem[\protect\citeauthoryear{{Begelman}, {Blandford} \& {Rees}}{{Begelman}
  et~al.}{1980}]{begelman80}
{Begelman} M.~C.,  {Blandford} R.~D.,    {Rees} M.~J.,  1980, \nat, 287, 307

\bibitem[\protect\citeauthoryear{{Berczik}, {Merritt} \& {Spurzem}}{{Berczik}
  et~al.}{2005}]{berczik05}
{Berczik} P.,  {Merritt} D.,    {Spurzem} R.,  2005, \apj, 633, 680,
  astro-ph/0507260

\bibitem[\protect\citeauthoryear{{Berczik}, {Merritt}, {Spurzem} \&
  {Bischof}}{{Berczik} et~al.}{2006}]{berczik06}
{Berczik} P.,  {Merritt} D.,  {Spurzem} R.,    {Bischof} H.-P.,  2006, \apjl,
  642, L21, astro-ph/0601698

\bibitem[\protect\citeauthoryear{{Gopal-Krishna,}, {Biermann} \&
  {Wiita}}{{Gopal-Krishna,} et~al.}{2003}]{gopal-krishna03}
{Gopal-Krishna,} {Biermann} P.~L.,    {Wiita} P.~J.,  2003, \apjl, 594, L103,
  astro-ph/0308059

\bibitem[\protect\citeauthoryear{{Heggie}}{{Heggie}}{1975}]{heggie75}
{Heggie} D.~C.,  1975, \mnras, 173, 729

\bibitem[\protect\citeauthoryear{{Hills}}{{Hills}}{1975}]{hills75a}
{Hills} J.~G.,  1975, \aj, 80, 809

\bibitem[\protect\citeauthoryear{{Holley-Bockelmann}, {Mihos}, {Sigurdsson},
  {Hernquist} \& {Norman}}{{Holley-Bockelmann}
  et~al.}{2002}]{holley-bockelmann02}
{Holley-Bockelmann} K.,  {Mihos} J.~C.,  {Sigurdsson} S.,  {Hernquist} L.,
  {Norman} C.,  2002, \apj, 567, 817, astro-ph/0111029

\bibitem[\protect\citeauthoryear{{Holley-Bockelmann} \&
  {Sigurdsson}}{{Holley-Bockelmann} \&
  {Sigurdsson}}{2006}]{holley-bockelmann06}
{Holley-Bockelmann} K.,  {Sigurdsson} S.,  2006, astro-ph/0601520

\bibitem[\protect\citeauthoryear{{Hut}}{{Hut}}{1983}]{hut83}
{Hut} P.,  1983, \apjl, 272, L29

\bibitem[\protect\citeauthoryear{{Merritt} \& {Ekers}}{{Merritt} \&
  {Ekers}}{2002}]{merritt02}
{Merritt} D.,  {Ekers} R.~D.,  2002, Science, 297, 1310, astro-ph/0208001

\bibitem[\protect\citeauthoryear{{Milosavljevi{\' c}} \&
  {Merritt}}{{Milosavljevi{\' c}} \& {Merritt}}{2001}]{milos01}
{Milosavljevi{\' c}} M.~.,  {Merritt} D.,  2001, \apj, 563, 34,
  astro-ph/0103350

\bibitem[\protect\citeauthoryear{{Milosavljevi{\' c}}, {Merritt}, {Rest} \&
  {van den Bosch}}{{Milosavljevi{\' c}} et~al.}{2002}]{milos02}
{Milosavljevi{\' c}} M.,  {Merritt} D.,  {Rest} A.,    {van den Bosch} F.~C.,
  2002, \mnras, 331, L51, astro-ph/0110185

\bibitem[\protect\citeauthoryear{{Peters}}{{Peters}}{1964}]{peters64}
{Peters} P.~C.,  1964, \prb, 136, 1224

\bibitem[\protect\citeauthoryear{{Quinlan}}{{Quinlan}}{1996}]{quinlan96}
{Quinlan} G.~D.,  1996, New Astronomy, 1, 35, astro-ph/9601092

\bibitem[\protect\citeauthoryear{{Quinlan} \& {Hernquist}}{{Quinlan} \&
  {Hernquist}}{1997}]{quinlan97}
{Quinlan} G.~D.,  {Hernquist} L.,  1997, New Astronomy, 2, 533,
  astro-ph/9706298

\bibitem[\protect\citeauthoryear{{Rauch} \& {Tremaine}}{{Rauch} \&
  {Tremaine}}{1996}]{rauch96}
{Rauch} K.~P.,  {Tremaine} S.,  1996, New Astronomy, 1, 149, astro-ph/9603018

\bibitem[\protect\citeauthoryear{{Rottmann}}{{Rottmann}}{2001}]{rottmann01}
{Rottmann} H.,  2001, PhD thesis, University of Bonn

\bibitem[\protect\citeauthoryear{{Saslaw}, {Valtonen} \& {Aarseth}}{{Saslaw}
  et~al.}{1974}]{saslaw74}
{Saslaw} W.~C.,  {Valtonen} M.~J.,    {Aarseth} S.~J.,  1974, \apj, 190, 253

\bibitem[\protect\citeauthoryear{{Toomre} \& {Toomre}}{{Toomre} \&
  {Toomre}}{1972}]{toomre72}
{Toomre} A.,  {Toomre} J.,  1972, \apj, 178, 623

\bibitem[\protect\citeauthoryear{{Yu}}{{Yu}}{2002}]{yu02}
{Yu} Q.,  2002, \mnras, 331, 935, astro-ph/0109530

\bibitem[\protect\citeauthoryear{{Zier}}{{Zier}}{2005}]{zier05}
{Zier} C.,  2005, \mnras, 364, 583, astro-ph/0507129

\bibitem[\protect\citeauthoryear{{Zier} \& {Biermann}}{{Zier} \&
  {Biermann}}{2001}]{zier01}
{Zier} C.,  {Biermann} P.~L.,  2001, \aap, 377, 23, astro-ph/0106419

\bibitem[\protect\citeauthoryear{{Zier} \& {Biermann}}{{Zier} \&
  {Biermann}}{2002}]{zier02}
{Zier} C.,  {Biermann} P.~L.,  2002, \aap, 396, 91, astro-ph/0203359

\end{thebibliography}
\bibliographystyle{my-mn2e}

\label{lastpage}

\end{document}